\documentclass{article}

\usepackage{conf_iap}
\usepackage{graphics}
\usepackage{psfig}

\newcommand{\cm}{\rm cm} 
\newcommand{\s}{{\rm s}}

\newcommand{\K}{{\rm K}}
\newcommand{\kpc}{{\rm kpc}}

\newcommand{\lya}{Ly$\alpha$} 

\begin{document}

\heading{Explaining the Lyman-alpha forest}
\author{Joop Schaye$^{1}$}
\address{Institute for Advanced Study, Princeton NJ 08540}

\begin{abstract}
It is shown that many properties of \lya\ forest absorbers can be
derived using simple physical arguments. Analytical expressions are
derived for the density and the size of an absorber as a function of
its neutral hydrogen column density, which agree well with both
observations and hydrodynamical simulations. An expression is
presented to compute $\Omega_{\rm IGM}$ from the observed column
density distribution, independent of the overall shape of the
absorbers. Application to the observed column density distribution
shows that at high redshift most of the baryons are in the forest
and suggests a simple interpretation for its shape and evolution.
\end{abstract}

\section{Introduction}
In the last decade semi-analytic models (e.g., \cite{bi92:lya}) and
hydrodynamical simulations (see \cite{efstathiou00:lya} for a recent
review) have been used to show that cold dark 
matter models with a nearly scale-invariant spectrum of initial,
adiabatic, Gaussian fluctuations are very successful in reproducing
the large body of high-quality observations of the \lya\ forest. The
physical picture that has emerged is that the forest arises in a
network of sheets, filaments, and halos, which give rise to absorption
lines of progressively higher column densities. The low column density
absorption lines ($N_{HI} \lsim 10^{14.5}~\cm^{-2}$ at $z\sim 3$), arise
in a smoothly fluctuating intergalactic medium (IGM) of moderate
overdensity ($\rho < 10\left < \rho \right >$), and contain most of
the baryons in the universe. On large scales the gas traces the dark
matter and observations of the \lya\ forest can be used to study the
large-scale distribution of matter, while on small scales
hydrodynamical effects are important and the detailed line profiles
can be used to reconstruct the thermal history of the IGM.
In this contribution I will show that the modern picture of the forest
can be derived directly from the observations using straightforward
physical arguments, without making any assumptions about the presence
of dark matter, the mechanism for structure formation or the precise
cosmological model. This work is described in more detail in a
recent publication \cite{schaye01:lya}.

\section{Physics}
Consider an (optically thin) self-gravitating gas cloud of arbitrary
shape that is intersected by our line of sight to a background
QSO. Let $n_H$ be the characteristic total hydrogen number density of
the absorber, i.e., the density weighted by the neutral hydrogen
column density and let $L$ be the characteristic size of the absorber,
i.e., the size over which the density is of order $n_H$.

Regardless of whether the cloud as a whole is in dynamical
equilibrium, it will in general not be far from local hydrostatic
equilibrium, i.e., $L \sim L_J$ where $L_J$ is the local Jeans
length. If $L \ll L_J$, then the cloud will expand or evaporate and
equilibrium will be restored on a sound-crossing time scale. If $L \gg
L_J$, then the cloud is Jeans unstable and will fragment or shock and
equilibrium will be restored on a dynamical time scale. Note that this
argument breaks down if the relevant time scale is larger than the
Hubble time, because in that case there simply has not been sufficient
time to establish equilibrium. Since $t_{\rm dyn} \equiv (G\rho)^{-1/2}$
and $t_H \equiv H^{-1} \sim (G\left <\rho\right >)^{-1/2}$ this is true for
underdense absorbers. Such absorbers have sizes greater than the
sound horizon and can be regarded as a ``fluctuating Gunn-Peterson
effect''. Systematic departures from local hydrostatic
equilibrium ($L \ll\gg L_J$) also occur in clouds confined by external
pressure (but the large cloud sizes derived from observations of quasar
pairs rule out the possibility that a significant fraction of the forest
arises in pressure confined clouds \cite{rauch98:lya}), and in
rotationally supported clouds whose spin axis is nearly perpendicular
to our sight line.

\section{Results}
Combining the condition of local hydrostatic equilibrium, $L\sim L_J$,
with the expression for the neutral fraction in an optically thin
plasma, $n_{HI}/n_H \approx 0.46 n_H T_4^{-0.76} \Gamma_{12}^{-1}$
(where $T_4 \equiv T / 10^4~\K$ is the temperature and $\Gamma_{12}
\equiv \Gamma / 10^{-12}~\s^{-1}$ is the hydrogen photoionization
rate), yields expressions for the density and size of an absorber as a 
function of its neutral hydrogen column density \cite{schaye01:lya}:
\begin{eqnarray}
N_{HI}& \sim & 2.7 \times 10^{13} ~\cm^{-2} ~ (\rho / \left < \rho
\right >)^{3/2}
T_4^{-0.26} \Gamma_{12}^{-1}
\left ({1+z \over 4}\right )^{9/2}
\left ({\Omega_b h^2 \over 0.02}\right )^{3/2}
\left ({f_g \over 0.16} \right )^{1/2}, \label{eq:NHI} \\
L &\sim& ~ 1.0\times 10^2 ~\kpc ~ 
\left ({N_{HI} \over 10^{14}~\cm^{-2}}\right )^{-1/3}
T_4^{0.41}
\Gamma_{12}^{-1/3}
\left ({f_g \over 0.16}\right )^{2/3},
\label{eq:L}
\end{eqnarray}
where $f_g \approx \Omega_b / \Omega_m$ is the fraction of the mass in 
gas (excluding stars and 
molecules). From the line widths of the absorption lines it is
known that the temperature is roughly consistent with that
expected from photoionization heating, $T_4\sim 1$. Studies of the
proximity effect estimate $\Gamma_{12} \sim 1$ at $z\sim 3$ and
$\Gamma_{12} \sim 10^{-1}$ at $z\sim 0$ (e.g.,
\cite{scott01:proximity}).

The density - column density relation agrees with published results
from hydrodynamical
simulations to within a factor of a few, about as well as different
simulations agree with each other. The sizes of the absorbers have not
been investigated in detail with simulations, but equation \ref{eq:L}
does agree with the sizes derived from observations of multiple sight
lines (e.g., \cite{bechtold94:lya}). 

Using equation \ref{eq:NHI} to compute the neutral fraction as a
function of column density, we can compute $\Omega_{\rm IGM}$, the
contribution of the \lya\ forest to the cosmic baryon density,
directly from the observed column density distribution: 
\begin{equation}
\Omega_{\rm IGM} \sim 2.2 \times 10^{-9} h^{-1}
\Gamma_{12}^{1/3}
\left ({f_g \over 0.16}\right )^{1/3}
T_4^{0.59} {H(z) \over H_0}{1 \over (1+z)^2} 
\int N_{HI}^{1/3}  {d^2 n \over dN_{HI}dz} \,dN_{HI}.
\label{eq:omegab}
\end{equation}
Figure 1a shows the observed column density distribution at $z\sim
3$ and the single power law fit of Hu et al.\ (1995) (dashed line).
Integration of this column density distribution according to equation
\ref{eq:omegab}, using
$(T_4,\Gamma_{12},f_g,h,\Omega_m,\Omega_\Lambda)=(2.0,1.0,0.16,0.65,0.3,0.7)$,
yields $\Omega_{\rm IGM}h^2 \approx 0.015$ for $\log N_{HI} = 13.0 -
17.2$, confirming that at high $z$ most of the baryons are in the
diffuse IGM.

Figure 1b shows the derived mass distribution, i.e., $d\Omega_{\rm
IGM}/d\log(1+\delta)$. The shape of the mass distribution clearly
reflects the deviations from a single power law column density
distribution and agrees reasonably well with the actual mass
distribution in a hydrodynamical simulation (solid line), kindly
provided by T.~Theuns. The agreement becomes even more impressive if
we compare with the distribution of gas that has a temperature $\log T
< 4.8$ (dashed line), low enough for collisional
ionization to be ineffective. This good agreement suggests that the
observed structure in the IGM formed through gravitational instability
in an expanding universe, as is the case in the simulation.

The shape of the mass distribution derived from the observations also
suggests a simple physical interpretation for the deviations from the
single power law in the observed column density distribution, whose
origin was not yet understood. The steepening at $N_{HI} \sim
10^{14.5}~\cm^{-2}$ reflects the fall-off in the density distribution
due to the onset of rapid, non-linear collapse at $\delta\sim
10$. Finally, the flattening at $N_{HI} \sim 10^{16}~\cm^{-2}$ can be
attributed to the flattening of the density distribution at $\delta
\sim 10^2$ due to the virialization of collapsed matter.

The evolution of the shape of
the column density distribution can now also be explained. 
From $z\sim 3$ to $z\sim 0$, the slope of the column density
distribution is found to steepen for $N_{HI} 
\sim 10^{13}$--$10^{14}~\cm^{-2}$ and to flatten for $N_{HI} 
\sim 10^{15}$--$10^{16}~\cm^{-2}$
(e.g., \cite{penton00:lya}). If gravitational instability is
responsible for the shape of the gas density distribution, then this
distribution will be similar at $z\sim 0$ and $z\sim 3$, provided it
is expressed as a function of the density relative to the
cosmic mean. Hence, the change in the slope of the column density
distribution must mainly be due to the evolution of the function
$N_{HI}(\delta)$, which is determined by the expansion of the universe and the
evolution of the ionizing background. Equation \ref{eq:NHI} shows
that $N_{HI}\propto (1+z)^{9/2}/\Gamma$. The neutral hydrogen 
column density corresponding to a fixed density contrast is thus about a
factor $5\times 10^2\,\Gamma(z=0)/\Gamma(z=3) \sim 5$--50 lower at
$z\sim 0$ than at $z\sim 3$, and this can account for the evolution in
the observed distribution.

This work demonstrates that the main aspects of the physical picture
of the \lya\ forest, including the facts that the absorbers are
extended, of low overdensity, and contain a large fraction
of the baryons, can all be derived directly from the observations
using straightforward analytical arguments, without making
assumptions about the presence of dark matter, the mechanism for
structure formation, or the precise cosmology. Although this analysis
is essentially model-independent, the shape of the derived mass
distribution strongly suggests that gravitational instability in an
expanding universe is responsible for the distribution of matter in
the IGM, thus lending credence to the use of ab-initio
hydrodynamical simulations to investigate the detailed properties of
the forest and their dependence on the parameters of the models.

\begin{figure}
\centerline{\hbox{
\psfig{figure=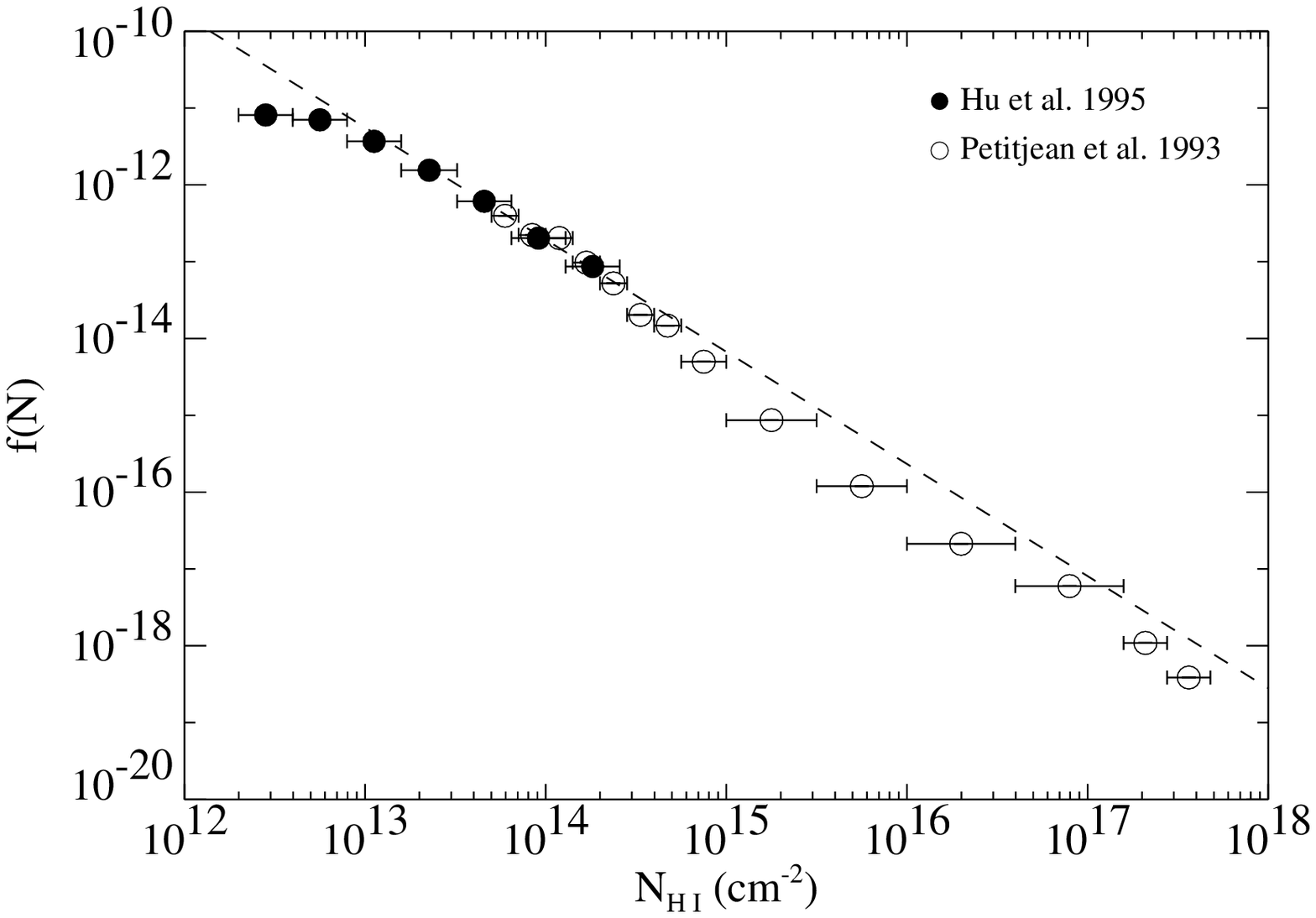,height=0.45\textwidth}
\psfig{figure=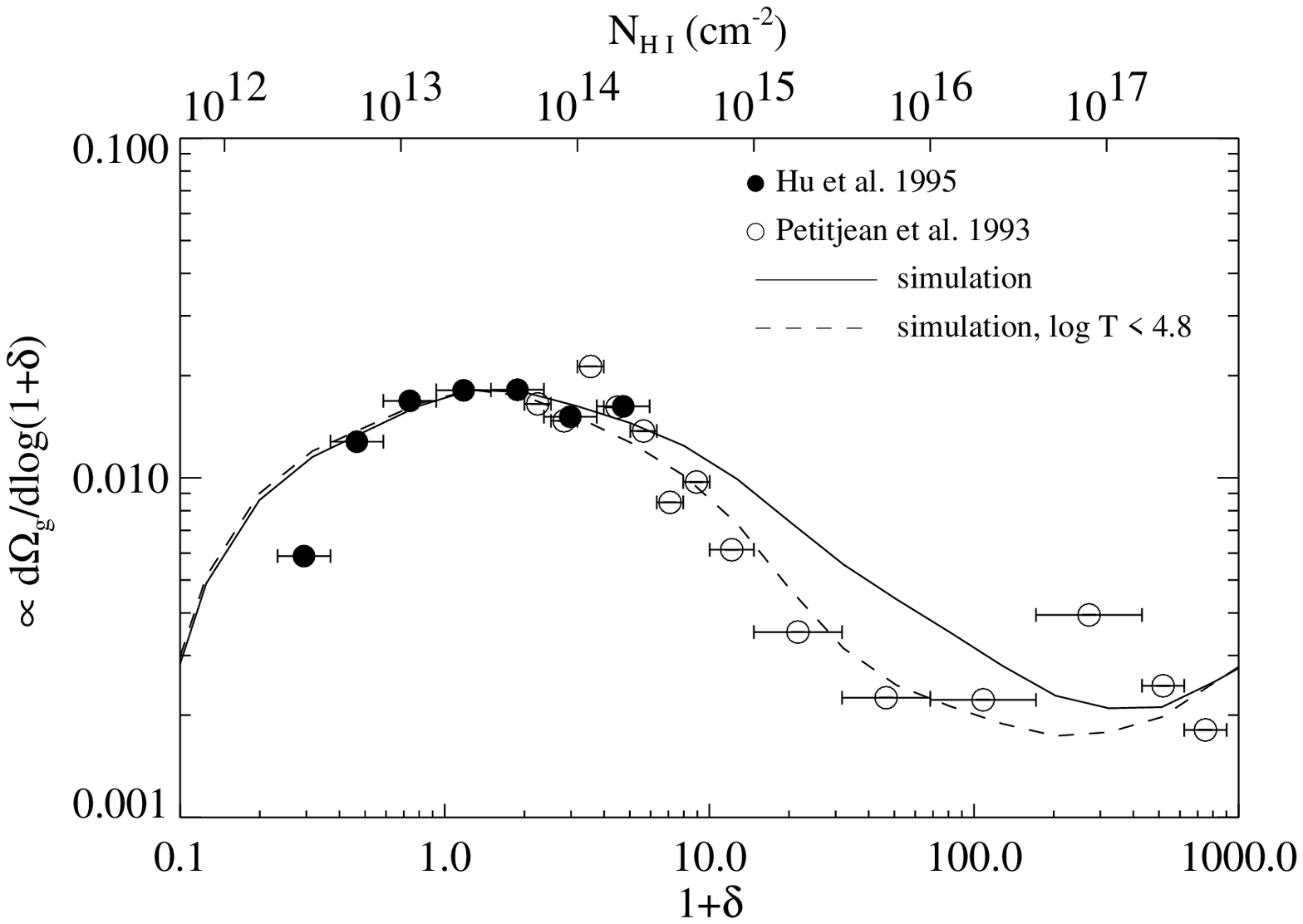,height=0.45\textwidth}}}
\caption[]{(a) Observed column density distribution function at $z\sim
3$. (b) The mass distribution as a function of overdensity, derived form the
observed column density distribution using equation 4 and
$(T_4,\Gamma_{12},f_g,h,\Omega_m,\Omega_\Lambda)=(2.0,1.0,0.16,0.65,0.3,0.7)$.
The curves, which have been
normalized to the maximum of the Hu et al.\ data points, indicate the
mass distribution in a hydrodynamical simulation with the same
cosmological parameters as were used to calculate the data points.}
\end{figure}

\end{document}